\documentclass[12pt]{article}
\begin{document}
\pagestyle{empty}
\begin{center}  {\Large \bf Possible Experiments to test Einstein's Special Relativity Theory  }
\\~\\ {\normalsize  Victor-Otto de Haan} ~ \\
{\it (c) 2011 BonPhysics, Laan van Heemstede 38, 3297 AJ Puttershoek, The Netherlands} \end{center} ~ \\
{\bf Abstract.} All of the experiments supporting Einstein's Special Relativity Theory are also supportive of the Lorentz ether theory, or many other ether theories. However, a growing number of experiments show deviations from Einstein's Special Relativity Theory, but are supporting more extended theories. Some of these experiments are reviewed and analyzed. Unfortunately, many experiments are not of high quality, never repeated and mostly both. It is proposed that the most promising experiments should be repeated. 

\begin{center}{\large \bf Introduction}\end{center}
Experiments using light interference, electromagnetic phenomena or high energy elementary particles are all supportive for Einstein's Special Relativity Theory. In general it is not known that these experiments can be equally well explained by ether theories. For example it was shown by Lorentz~\cite{Lorentz1915} that his earlier ether theory based on the Maxwell equations extended with point-particles and the Lorentz force is formally the same. The question automatically arises under which conditions an ether theory and special relativity might give different results. A possible answer has already been given by Helmholtz in 1858~\cite{Hemlholtz1858} and was based on a remark by Euler in a publication of 1755~\cite{Euler1755}. This was made to draw attention to limitations of velocity potentials to describe fluid motion. Triggered by this, Helmholtz showed that some type of fluid motion can not be generated or destroyed by conservative forces. He related this kind of fluid motion to vortex motion. Santilli~\cite{Santilli1} discovered a similar shortcoming in the modern use of Lagrangian and Hamiltonian dynamics, where standard all non-conservative forces are neglected. Santilli~\cite{Santilli1} describes two conditions under which these neglects are not detrimental. The first one is the {\it closure condition}: The system can be considered as isolated from the rest of the universe in order to permit the conservation laws of the total mechanical energy, the total physical linear momentum, the total physical angular momentum, and the uniform motion of the center of mass. The second one is the {\it selfadjointness condition}: The particles can be well approximated as massive points moving in vacuum along stable orbits without collisions, in order to restrict all possible forces to those of action-at-a-distance, potential type. Hence, the search for experiments to invalidate special relativity should be based on considerations whether or not these experiments violate these conditions. Todays examples of such experiments are (among many others) the IsoShift experiments~\cite{SantilliIsoRedShift}, the superluminal tunneling experiments~\cite{Nimtz} and the rotational M\"{o}ssbauer experiments~\cite{RotMoess}.

Here, the focus is on experiments which might give evidence for the existence of an absolute reference frame in which the ether is at rest. It is clear from the preceding that experiments are needed which violate either or both of the conditions referenced above.

\begin{center}{\large \bf Classification of experiments}\end{center}

Simply one can divide the experiments to determine the absolute motion of the reference frame (or in other terms 'of the ether') into two categories: first order or second order experiments, where the observed effect should be proportional to the appropriate order of the ratio of the velocity of the laboratory frame relative to the speed of light. 

Bradley aberration~\cite{Bradley} and the cosmic microwave background signal~\cite{Smooth1977} are the most famous ones of the first category, but these are already interpreted differently by mainstream physics. The observation of a dipole distribution in the cosmic microwave background radiation~\cite{Smooth1977} is an important experiment. By special relativity fans it is interpreted as the remnants of the initiation of the universe. For others it is a clear indication of a preferred reference frame and for some it has triggered renewed interest in the old ether concept. If it is interpreted as the frame in which the ether is at rest, another conclusions must be drawn from the observation of the dipole: A first order effect is possible. This is in direct contrast to the popular believes of the 20th century.

The Michelson Morley experiment~\cite{Michelson1881} is the most famous one for the second category. Because of the large speed involved and the smallness of velocity of the laboratory, in the 19th and first half of the 20th century, measurements were restricted to interference techniques (polarisation measurement can also be interpreted as an interference technique). The attention changed from first order experiments to second order experiments when at the end of the 19th century the Fizeau drag effect was used to explain why first order experiments were not able to detect the absolute speed of the earth. Nowadays, a further distinction into two other categories can be made: interference measurements and non-interference experiments. 

In table~\ref{tab1} the categories with some examples are shown. Some of these experiments have been performed, but never repeated. Others are proposals based on theoretical analysis. The listing is typical, but incomplete. In the following, first the history of the second order interference experiment is discussed. This type of experiment is not a perfect candidate for a possible experiment to find deviations from special relativity because of the smallness of the effect: they are second order effects and the deviations from the Santilli conditions are also small. 

First order interference experiments as proposed by M\'{u}nera~\cite{MuneraPriv}, Spaveri~\cite{Spaveri} and Wesley~\cite{Wesley} and performed by Silvertooth~\cite{Silvertooth,Silvertooth1} and De~Haan~\cite{deHaan2010} are therefore better candidates. It is argued that time-of-flight measurements might do a better job as claimed via experiments by Marinov~\cite{Marinov2007} and De~Witte~\cite{DeWitte} and proposed by Kozynchenko~\cite{Kozynchenko2006} and Sardin~\cite{Sardin} and in progress by Ahmed~\cite{Ahmed2011}. Finally, a completely novel possibility that is proposed by Christov~\cite{ChristovCor} is discussed.

\begin{table}[bhtp] \begin{tabular}{|l|l|l|} 
\hline
& \ Experiment & \ Proposal \\ \hline
\begin{tabular}{l}Interference \\First order \end{tabular} 
&
\begin{tabular}{l} Silvertooth (Standing waves) \\ Galaev (Dynamic) \\ De Haan (Gas-filled) \end{tabular}
&
\begin{tabular}{l} Wesley (Adapted Sagnac) \\ Spaveri (Material-filled) \\ Munera (Gas-filled) \\ Christov (Correlator) \end{tabular} 
\\ \hline
\begin{tabular}{l} Interference \\ Second order \end{tabular} 
&
\begin{tabular}{l} Michelson Morley  \\ Demjanov (Material-filled)  \\ Munera (Stationary) \\ Cahill (Optical fiber) \\ De Haan (Optical fiber) \end{tabular} 
&
\begin{tabular}{l} Consoli (Gas-filled)  \end{tabular} 
\\ \hline
\begin{tabular}{l} Non-Interference \\ First order \end{tabular} 
&
\begin{tabular}{l} Bradley aberration \\ Cosmic Microwave Background \\ Marinov (Coupled shutters) \\ De Witte (Time difference)  \end{tabular} 
&
\begin{tabular}{l} Ahmed (Coupled shutters) \\ Kozynchenko (Time diff.) \end{tabular} 
\\ \hline
\begin{tabular}{l} Non-Interference \\ Second order \end{tabular} 
&
&
 \begin{tabular}{l} Sardin (Time difference) \\ Phipps, Jr. (Bradley aber.) \end{tabular}  
\\ \hline 
\end{tabular}
\caption{Categories and possible experiments to test special relativity theory}
\label{tab1} \end{table}

\begin{center}{\large \bf Second order interference experiments}\end{center}

In 1881 Michelson~\cite{Michelson1881} devised an apparatus that should be able to measure the change of the velocity very accurately. The apparatus is now known as a Michelson Morley interferometer. After some comments on the experiment by Lorentz in 1886~\cite{Lorentz1886} Michelson and Morley~\cite{Michelson1887} increased the sensitivity of the apparatus with almost a factor of ten overcoming the accuracy objections of Lorentz. The accuracy of the apparatus was further increased with a factor of 6 by Morley and Miller~\cite{Morley1905} and by Miller in a series of experiments between 1905 and 1930~\cite{Miller1922,Miller1926,Miller1930,Miller1933}. In all these experiments the sought for magnitude of the effect was never observed. However, Miller in his elaborate series of experiments, always claimed that he measured a small second order effect and also a first order effect. The second order effects he measured were quite small with respect to the sought for effect, but larger than the experimental error. These second order effects were analysed by him by combining measurements at different epochs. Combining the results from these epochs and assuming the Sun moves relative to the preferred rest frame he was able to find a preferred direction in space and a velocity. The first order effect he measured depended very much on the detailed experimental settings and were not analysed to find an anisotropy. In view of this discrepancy several researchers try to find experimental evidence of first or second order effects with Michelson Morley interferometer type instruments. This has been done by, for instance, Piccard~\cite{Piccard1926,Piccard1928}, Illingworth~\cite{Illingworth1927} and Joos~\cite{Joos1930}. All these authors report the absence of the sought for effect. However, according to M\'{u}nera~\cite{Munera1998,Munera2006} these experiments all have results comparable with those of Miller. Hence, experimental evidence is not conclusive whether or not some first or second order effect exists. Recently, it has been argued by Cahill~\cite{Cahill2003} and Consoli~\cite{Consoli2004} that the Miller effect~\cite{Miller1933}, together with all other Michelson Morley interferometer experiment results~\cite{Piccard1926,Piccard1928,Illingworth1927,Joos1930}, could be caused by a reduction of ether drag. This drag would depend on the difference of the refractive index of 1, which for atmospheric air is approximately $3 \times 10^{-4}$, for atmospheric helium $4 \times 10^{-5}$ and for vacuum $0$. This would also explain why modern-day vacuum experiments all give much lower limits for the anisotropy. Experiments performed by Demjanov~\cite{Demjanov2010} and Galaev~\cite{Galaev2002} seem to confirm these predictions, but they have never been repeated. Cahill~\cite{Cahill2008} used a fiber optic interferometer and claimed a positive result. This experiment was repeated by De~Haan~\cite{DeHaan2009} under (almost) the same conditions yielding a result compatible with special relativity. The idea that the drag would depend on the medium (or is time-dependent as assumed by Galaev) can also be explained by a violation of one of the Santilli's conditions mentioned earlier. Hence, there could be a relation between the Santilli IsoShift~\cite{SantilliIsoRedShift} and the Michelson Morley anomolies.

\begin{center}{\large \bf First order interference experiments}\end{center}

Due to the smallness of second order effects many have devised experiments that should give a first order effect. Successful candidates are experiments which incorporate a violation of Santilli's conditions. This could be due the interaction of light with matter as discussed in the previous section. 

According to M\'{u}nera~\cite{MuneraPriv} and Spaveri~\cite{Spaveri} the second order effects mentioned could be transformed into a first order effect by using an a-symmetric Mach-Zehnder interferometer. One arm of the interferometer contains over a path length $L$ a material with refractive index $n_1$ and the other arm over the same length a material with refractive index $n_2$. Spaveri calculates a change in traveling time difference in the two arms upon rotation of the setup of $\Delta t=2vL(n_1^2-n_2^2)/c^2$ where $v$ is the velocity of the ether wind, $c$ the velocity of light. He then argues that this will yield a fringe shift proportional to first order and would result in an easy obtained detection limit for $v$ of some meters per second. A fiber optical version of this experiment was performed by De~Haan~\cite{deHaan2010}. In one arm a glass tube was inserted with a length of 100 mm that could be filled with atmospheric air or helium. When the glass tube was filled with air, upon rotation a fringe shift was observed corresponding to a maximum velocity of 64(6)~km/s, about twice the velocity of Earth in its orbit around the Sun. However, the azimuth of the maximum of the first order effect was in the North to South direction and did not depend on sidereal time. When the air was replaced by helium this shift remained almost the same, casting doubts on the validity of Consoli's assumption of the reduction of ether drag. Another possible explanation would be that the ether velocity is dependent by the height above the surface of the Earth. This effect is mentioned by Miller~\cite{Miller1933} as a possible explanation for his reduced effect. Galaev~\cite{Galaev2002} introduces such an effect to explain his measurements results with an s-symmetic Mach-Zehnder interferometer. Such an effect could also depent on the medium surrounding the experiment, for it is not known to what extension the ponderable matter might influence the ether velocity. For definite conclusions these experiment need to be repeated with higher accuracy and at several altitudes.

Wesley~\cite{Wesley} describes an interesting possibility that (as far as the author is aware) has never been performed. He uses a Mach-Zehnder type of interferometer and analyses the resulting intensities of independent beams passing in opposite directions through the interferometer in a frame that is both rotating and translating. The novelty is in the comparison of intensities produced by counter propagating waves at two different locations. There might be a connection to the experiment performed by Silvertooth~\cite{Silvertooth,Silvertooth1}. He used a very thin transparent photo detector~\cite{Silvertooth2} to detect the nodes of the standing wave created by two counter propagating waves in a Sagnac type of interferometer. Silvertooth claimed a positive result but the theoretical background of the experiment was never explained satisfactorily~\cite{STExp0,STExp1,STExp2,STExp3,STExp4,STExp5}. The experiment was repeated by Marinov twice. First with a similar result~\cite{Marinov1987} and later after adaptation of the experiment with a negative result~\cite{Marinov1988}. The adaptation was the replacement of the standing wave detector by a transparent mirror, changing the interference from counter propagating waves into interference of waves traveling into the same direction. This indicates that the use of counter propagating waves is crucial. The connection between Wesley's proposal and Silvertooth experiment can be made by the Wang's description of a Generalized Sagnac effect~\cite{Wang2004} as due to any moving part of the experiment. In Silvertooth experiment, the rotation of the earth would be used to create the rotational motion additional to the translation of the solar system. The possible accuracy of Silvertooth experiment makes it a very attractive option to reproduce.

\begin{center}{\large \bf Second order non-interference experiments}\end{center}

Interference techniques are regarded as the most accurate ones for the detection of the preferred frame. However, standard interference techniques use interference between light waves traveling in the same direction to obtain intensity fluctuations or fringes due to travel distance differences and not due to travel time differences. If the wave character of light is taken into account, light reflected from a moving mirror obtains in general a different frequency. If the Doppler effect is taken into account, this complicates the calculations. Further complications arise due to the aberration effect. Under these conditions it might be considered that Lorentz contraction and/or time dilatation does not occur in reality. Based on this reasoning, Sardin~\cite{Sardin} proposed to measure the actual time difference of the travel time of the light beams through the two arms of a Michelson Morley interferometer. With current state-of-the-art pulsed lasers and an interferometer as large as LIGO with multiple reflections yielding an effective arm length of 120~km the expected time difference is some nanoseconds. It was considered not feasible by LIGO staff~\cite{SardinPriv}.

\begin{center}{\large \bf First order non-interference experiments}\end{center}

In 1728 Bradley~\cite{Bradley} discovered that some stars exhibited an aberration depending on the velocity of the earth around the sun. This is now known as Bradley aberration. Its explanation in the framework of special relativity is disputed in literature (see for instance~\cite{Gift2009}), especially since De~Sitter~\cite{DeSitter} showed that binary stars (moving with a different velocity at approximately the same location in the sky) have the same aberration independent of the velocity of the stars. The discrepancy can be mended up to first order if the wave character of light is taken into account. Phipps~\cite{Phipps2010} proposes that higher order terms might be observable by Very Long Base Line interferometry.

In the 1970-80's Marinov~\cite{Marinov1980,Marinov2007} performed several first order experiments which (he claimed) gave positive and similar results. They were all based on a so-called Newtonian time synchronisation. The idea that a Newtonian time synchronisation can be obtained is strengthened by the well-known clock paradox or twin paradox. It has been and still is discussed by many authors. It is closely related to the question whether time dilatation is a dynamical process or not. According to special relativity the observer's time scale is changed when he moves with respect to a clock. According to compatible ether theories the time scale is fixed (Newtonian) and the clock set in motion changes its rate. Based on the idea that Newtonian time synchronization is achieved, Kozynchenko~\cite{Kozynchenko2006} proposes to measure sidereal changes in the time-of-flight of laser pulses between two distant locations on the Earth surface. 

The Newtonian time synchronisation as realized by Marinov is based on shutters or mirrors mounted on two rotating discs connected by a rigid axis. Ives~\cite{Ives1939} showed that such a system, subjected to Lorentz contraction, cannot be used as a Newtonian time synchronisation. However, Lorentz contraction is based on conservative forces and it could be that contact forces in the axis and discs (violating the second Santilli condition) would enable a Newtonian time synchronisation. Again to the author knowledge Marinov's experiments have not been repeated until now, but an attempt is in progress~\cite{Ahmed2011}. 

In 1991 De~Witte~\cite{DeWitte} performed a first order experiment by measuring a time-phase delay of a 5~MHz electromagnetic signal through a 1.5~km long cable. The novelty of this experiment was that he did not use an interference technique to determine the time delay, but he directly measured the phase of the waves. He measured for 178 days and claimed to have observe a sidereal dependence on the occurrence of the maximum time delay. The experiment was never repeated in this way. Based on this experiment Cahill claimed to measure a similar effect~\cite{Cahill2006}, but it is based on very limited data. 

Christov~\cite{Christov2006} uses this idea of phase comparison in a novel way. Instead of measuring the local intensity of interfering counter propagating light beams he proposes to measure the correlation between the electromagnetic fields at different locations. The correlation between the electromagnetic fields should exhibit a clear first order effect, varying with the distance between the locations. The maximum effect occurs if the ratio between the wavelength of the used light and the distance between the locations is equal to the ration of the expected velocity and the velocity of light, i.e. 1/1000. For visible light the frequency is too high to be able to measure the temporal characteristics of the electromagnetic field. For lower frequencies down to radio waves this is possible. However, the associated wavelengths are much larger, which results in distances of the order of several meters to hundreds of meters to obtain accurate enough results. With the use of Terahertz waves (sub mm) the dimensions could be kept below 1~m. Another approach to measure the correlation between two distant locations, could be to use superluminal tunneling~\cite{Nimtz,Pereyra2007,Recami2009} or other non-propagating transfer mechanisms, like for instance a standing wave crossing an absorber~\cite{Cramer1980}.

\begin{center}{\large \bf Conclusions}\end{center}

It has been shown that a growing number of experimentalists are considering the possibility of detecting deviations from special relativity. 

To be able to experimentally test a theory a good understanding of its range of applicability is needed. An alternative theory that does not deviate in its experimental predictions can only be preferred or rejected by its meta-philosophical content. An alternative extended theory is needed to be able to device experiments to discriminate between them. The Maxwell-Lorentz ether theory extends special relativity (although it predates it too) as it uses absolute velocities, i.e. velocities relative to the frame in which the ether is at rest. However, as long as this extension is not experimentally verified is has no practical use and can be disregarded. 

Another extension has been realized by Santilli by incorporating contact forces or extended particles and non-locality. Contact forces can give rise to superluminal velocities which, when incorporated in a suitable experiment, should be able to expose the velocity of the ether. That is why in the above the considered experiments were focused on the detection of the ether rest frame. 

The above list is far from complete and only addresses certain experiments in which a possible violation of Santilli's conditions for the validity of special relativity is considered. Some experimentalists claim to have observed such a deviation. Unfortunately the reproduction of most of these experiments is either not documented or not performed. This omission clearly hinders scientific progress. On the one hand, if the reported deviations are experimentally confirmed, special relativity should have been replaced by a more extended theory. On the other hand, if they were experimentally dismissed, efforts could have been spent into other scientific endeavors. 

The most important experiments that needs to be reproduced are the first order experiments because of the expected magnitude of the effect. The interference measurements with counter propagating beams and a standing wave detector as performed by Silvertooth should be reproduced. The adapted Sagnac experiment as proposed by Wesley could be related to this experiment, however it does not use a standing wave detector so it is technically not too complicated. The novel experiment as proposed by Christov is interesting, not only to detect the ether rest frame, but also in studies where relative velocities are considered or when superluminal velocities are involved.

\end{document}